\documentclass[%
 reprint,
superscriptaddress,
 amsmath,amssymb,
 aps,
 prl,
]{revtex4-2}

\usepackage{graphicx}% Include figure files
\usepackage{dcolumn}% Align table columns on decimal point
\usepackage{bm}% bold math
\usepackage{subcaption,amsmath,gensymb,esvect,xcolor,xr,multirow,commath}
\usepackage[colorlinks=true,citecolor=blue]{hyperref}
\usepackage[normalem]{ulem}

\newcommand{\etal}{\textit{et al.}}
\newcommand{\mycomment}[1]{}

\begin{document}
\title{Surface Reconstructions in Thin-Films of Magnetic Topological Insulator MnBi$_2$Te$_4$}
\author{Shahid Sattar}
 \email{shahid.sattar@lnu.se}
  \affiliation{Department of Physics and Electrical Engineering, Linnaeus University, SE-39231 Kalmar, Sweden}
 \author{Daniel Hedman}
 \affiliation{Center for Multidimensional Carbon Materials (CMCM), Institute for Basic Science (IBS), Ulsan 44919, Republic of Korea}
 \author{C. M. Canali}
 \affiliation{Department of Physics and Electrical Engineering, Linnaeus University, SE-39231 Kalmar, Sweden}
 \date{\today}

\begin{abstract}
Understanding the nature of surface states and their exchange gaps in magnetic topological insulator MnBi$_2$Te$_4$ (MBT) thin films is crucial for achieving robust topological Chern and Axion insulating phases where the Quantum Anomalous Hall Effect and the Topological Magneto-electric Effect can be realized. Here, we focus on the rather unexplored issue of how surface reconstructions, which are likely to occur in experiments, influence these properties. Using first-principles calculations together with molecular dynamics simulations accelerated by machine learning force field, we demonstrate that interstitial-2H and peripheral-2H type atomic reconstructions are responsible for modifying the exchange gap and surface characteristics of MBT thin films, with important implications for the topological indices and the nature of quasi one-dimensional side-wall edge states dominating quantum transport. Specifically, the calculation of the energy landscape and barriers for the proposed surface reconstructions indicates that the interstitial-2H reconstruction is thermodynamically more stable than the peripheral-2H reconstruction. The latter case is also hypothesized as providing a plausible explanation for the Rashba surface states observed in Angle-Resolved Photoemission Spectroscopy (ARPES) measurements. Our analysis provides a theoretical framework to elucidate the nature and effect of reconstructions in MBT thin films, with predictions for the experimental realization of different topological phases.
\end{abstract}

\maketitle

%\section{Introduction}
\textbf{\textit{Introduction}} -- By merging the distinctive features of topological insulators (TIs) with time-reversal symmetry (TRS) breaking magnetism, the first intrinsic magnetic TI MnBi$_2$Te$_4$ (MBT) has undoubtedly opened an exciting new avenue in condensed matter physics aimed at unraveling new states of quantum matter and phenomena \cite{zhang2019topological,li2019intrinsic,otrokov2019unique}. This intricate combination of topology and magnetism has resulted in the experimental realization of the quantum anomalous Hall effect (QAHE) \cite{deng2020quantum,liu2020robust}, the Axion insulating phase \cite{liu2020robust,liu2021magnetic,lin2022direct}, Chern Anderson insulator \cite{li2024reentrant}, conductance fluctuations \cite{andersen2023universal}, and giant non-local edge conduction \cite{li2023giant}, while several other theoretical proposals are yet to be confirmed \cite{lian2020flat,li2024dissipationless,chen2024half}. Earlier work on magnetically doped TIs has addressed issues of film thickness dependence of the QAHE (such as in Cr-doped (Bi,Sb)$_2$Te$_3$) and modulation of the exchange gap with dopant concentration, asserting the need for homogeneous long-range magnetic ordering \cite{kou2012magnetically,chang2013thin,feng2016thickness}. These features and requirements are much better investigated and controlled in intrinsic magnetic TIs such as MBT, consisting of van der Waals (VdW)-coupled ferromagnetic (FM) septuple layers (SLs), with adjacent layers having opposite magnetization (see Fig.~\ref{fig:fig1}), leading to an overall antiferromagnetic (AFM) order down to the ultrathin few-layer limit. Understanding the rich and complex details of MBT not only advances fundamental physics but also holds potential for the use of magnetic TIs in spintronic devices and quantum computing technologies.

In contrast to magnetically doped TIs, MBT thin films display unique even-odd-layered dependent quantum effects, currently under intense investigation {\cite{zhao2021even,ovchinnikov2021intertwined,lupke2022local,li2024fabrication,chen2024even}, with unusual response to applied external magnetic and electric fields, field orientations, and temperature changes. Conservation of the combined $S=\Theta T_{1/2}$ symmetry in even-layered MBT films ($\Theta$: time-reversal operator, $T_{1/2}$: half-translation operator) ensures that each electronic energy band is doubly degenerate, in contrast to odd-layered MBT films where uncompensated magnetism results in the breaking of this symmetry and the consequent lifting of band degeneracy. Nevertheless, in both cases, a key quantity is the ``exchange gap" opened at the Dirac point of the surface states as a result of TRS breaking due to the FM exchange interaction between the localized Mn-atom magnetic moments within the same layer. Li \etal\cite{li2024imaging} have recently studied large fluctuations of the exchange gap (varying between 0 to 70 meV) in QAH or Chern insulating five-SL MBT samples, arguing that strong disorder in the magnetic texture is responsible for the gapless regions. Previously, Hou \etal\cite{hou2020te} showed that intrinsic Mn-Bi antisite defects and Te-vacancies at the surface of exfoliated few-layer MBT films lead to surface collapse and subsequent reconstruction, significantly affecting the surface electronic structure, as confirmed in later studies \cite{garnica2022native,wu2023irremovable}. These surface Mn-Bi antisite defects have also been implicated in the strong suppression of the exchange gap in the AFM phase \cite{HTan_BYan_PRL2023}. Furthermore, the observation of Rashba-like conduction bands further highlights the critical role of surface reconstructions \cite{vidal2019surface,nevola2020coexistence}. MBT thin films can exhibit both Chern and Axion insulating phases \cite{liu2020robust}, resulting in QHAE and yet-to-be verified topological magnetoelectric effect (TME), respectively. It is therefore essential to investigate theoretically unexplored and overlooked structural and surface modifications in MBT thin films that are responsible for the intricate bulk-surface correspondence from which these quantum effects originate.  

In this work, motivated by this complex experimental scenario, we have performed theoretical calculations on few-layer MBT thin films, and discovered that thermodynamically favorable interstitial-2H and peripheral-2H type reconstructions significantly affect topological properties such as the exchange gap and surface characteristics. Furthermore, we performed machine learning force field (MLFF) driven molecular dynamics (MD) at elevated temperatures to investigate the stability and transition between peripheral-2H and interstitial-2H types. Both types preserve the topological character of the unreconstructed system. However, for interstitial-2H the sidewall edge states are topological states leading to the QAHE and the Quantum Spin Hall Effect (QSHE). For peripheral-2H, additional non-topological Rashba states also emerge, corroborating angle-resolved photoemission spectroscopy (ARPES) measurements \cite{vidal2019surface,nevola2020coexistence}.  These results provide a theoretical framework to elucidate the fundamental properties of MBT thin films.

\textbf{\textit{Computational Methods}} -- First-principles calculations were performed using density functional theory (DFT) and the projector augmented wave method\,\cite{paw1,paw2} as implemented in the Vienna \textit{ab-initio} simulation package (VASP) \cite{vasp}. The generalized gradient approximation was used in the Perdew-Burke-Ernzerhof parametrization scheme to describe the effects of electronic exchange and correlation \cite{perdew1996generalized}. We used a plane wave cutoff energy of 450 eV \cite{perdew1996generalized}. The vdW dispersion energy corrections were incorporated using Grimme's DFT-D3 method \cite{grimme2010consistent}. For structural relaxations and self-consistent calculations (SCF), Brillouin zone integration was performed using a $\Gamma$-centered $9\times 9\times 1$ and $12\times 12\times 1$ K-mesh, respectively \cite{monkhorst1976special}. The atomic structures of the MBT thin films were relaxed until the residual forces acting on each atom were less than 10$^{-3}$ eV/\AA\, and an energy convergence criterion of 10$^{-6}$ eV was achieved. Additional details are given in the Supplementary Information (SI). 

\textbf{\textit{Results and Discussion}} -- Figure \ref{fig:fig1}(a-b) shows the schematic diagram of a five SL (5-SL) and six SL (6-SL) MBT thin film corresponding to the Chern and Axion insulating phases considered in our study. Each SL block represents seven atomic-plane thick MBT unit cell (Te-Bi-Te-Mn-Te-Bi-Te atoms)  having a homogeneous ferromagnetic alignment in $xy$-direction and A-type antiferromagnetic ordering in the $z$-direction -- see Fig.~S1 of the SI. Moreover, we find that the Mn magnetic moments in MBT have an out-of-plane easy axis, as indicated for each SL, which is consistent with earlier studies \cite{zhang2019topological,otrokov2019unique}. 

We start by describing the electronic band structure of {\it pristine}, i.e. relaxed but non-reconstructed odd- and even-layered thin films, shown in Fig.~\ref{fig:fig1}(c-d) for 5-SL and 6-SL thin films. We find that the surface exchange gaps opened at the Dirac point of the topological surface states are 53 meV and 59 meV, respectively.
Next we address the question of how individual MBT SLs and their constituent atomic-planes contribute to the exchange gap at the Dirac point. As shown in Fig.~S1 of the SI, the calculation of the partial density-of-states of the top topmost valence and bottom-most conduction bands projected on the entire atomic-plane spectrum clearly demonstrate that the major contribution to the electronic dispersion around the exchange gap comes from the atoms belonging to the very top and bottom SLs. 

\begin{figure}[!t]
\includegraphics[width=0.5\textwidth]{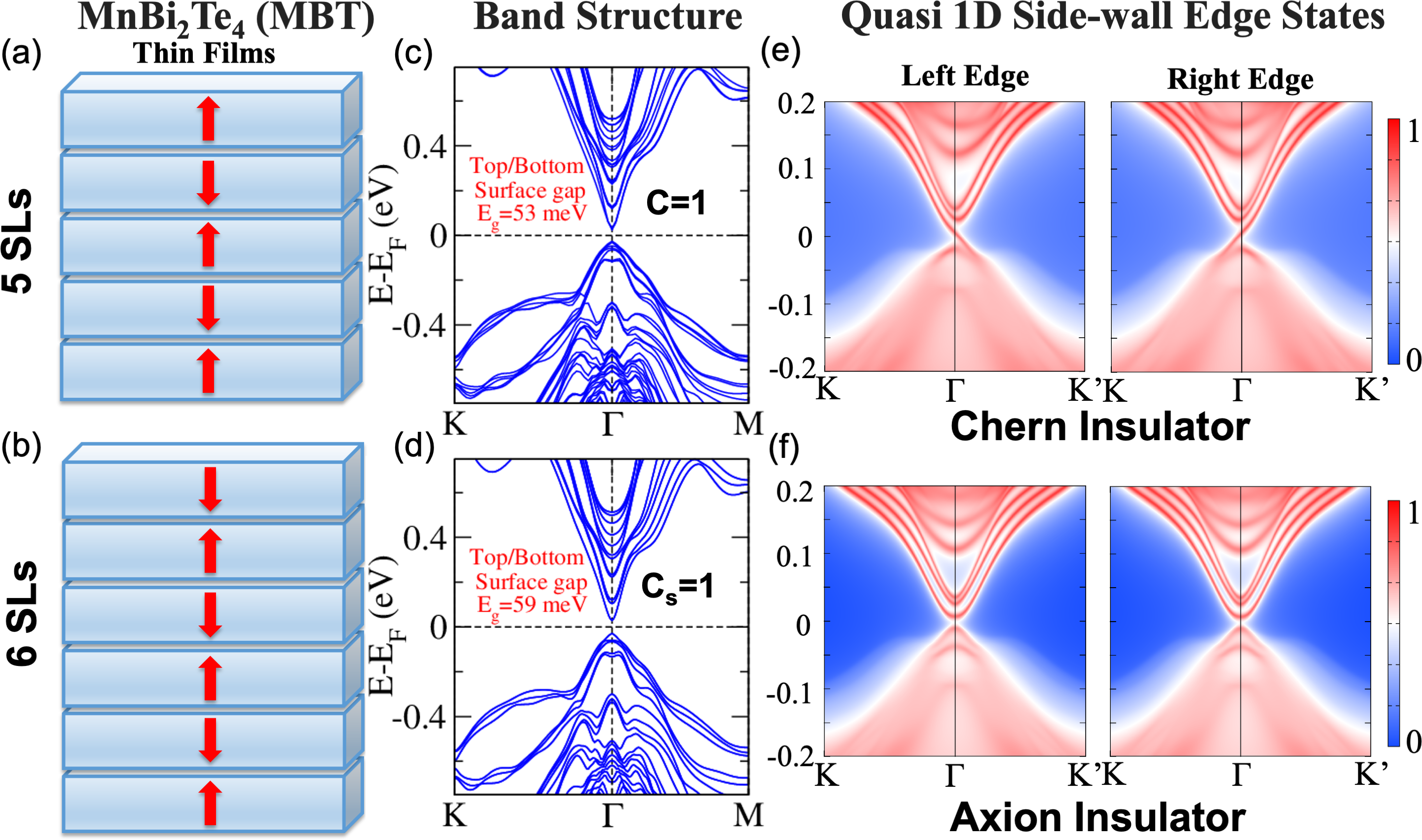}
\caption{(a,b) Schematic of a 5-SL and 6-SL MBT thin films having anti-parallel Mn spin orientation between neighboring SLs. (c,d) Electronic band structure, the magnitude of top/bottom surface energy gap, and topological invariants. (e,f) Quasi-1D left/right side-wall edge states of 110-surface, which is a characteristic feature of the Chern and Axion insulating phase.}
\label{fig:fig1}
\end{figure}

In order to confirm the topological nature of the MBT thin films, we next compute the Chern number ${\cal C}$ and the $Z_2$ topological invariants which take non-zero values for odd and even-layered systems, respectively. The Chern number (${\cal C} $) is obtained by taking the integral of the Berry curvature over a finite mesh of $K$-points spanning across the Brillouin zone by employing the Wannier Charge Center (WCC) method discussed in Ref. \cite{soluyanov2011computing}. We find that for the 5-SL MBT thin film ${\cal C} = 1$, confirming that this system is indeed a Chern insulator, whereas ${\cal C} = 0$ for the 6-SL film. Since inversion symmetry is absent in both the 5- and 6-SL MBT thin films, the 2D $Z_2$ topological invariant cannot be obtained by simply multiplying the parities of the occupied Bloch wave functions at the time-reversal invariant points in the Brillouin zone \cite{Fu2007}. Instead, we implement and use the computational method developed by Soluyanov \etal \cite{soluyanov2011computing} which can be employed also when inversion symmetry is absent. In this approach, $Z_2$ is defined in terms of a time-reversal polarization, implemented within a gauge corresponding to hybrid Wannier functions that are maximally localized in one dimension. Since a 6-SL possesses Kramer's degeneracy similar to bulk MnBi$_2$Te$_4$ and other AFM TIs \cite{mong2010antiferromagnetic}, $Z_2$ can be computed by tracking the evolution of WCC's across the Brillouin zone \cite{soluyanov2011computing}. For the 6-SL MBT thin film in the axion insulating phase, we find that the topological invariant $Z_2$ obtained by this method for this quasi-2D system is equal to $1$, implying that the system is a 2D topological insulator despite the lack of TRS. However, a classification  based on the Z$_2$ invariant and especially its evaluation in systems  where inversion symmetry and time-reversal symmetry are both absent (like a MBT thin film in the Axion insulator phase) is controversial. Therefore, we also considered the alternative approach of computing the spin Chern number (${\cal C}_s$) discussed in Refs.~\cite{prodan2009robustness,lin2024spin} using the computational approach proposed in Ref.~\cite{tyner2024berryeasy}. Indeed, Fig.~S2 in SI  shows the spectral flow of the spin-resolved WCC spectra and confirms the presence of non-trivial spin-Chern number (${\cal C}_s$=1) supporting the idea that the system is a quasi-2D topological insulator.

MBT thin films in either the Chern insulator or Axion insulator phase are characterized by 2D gapped top/bottom surface states due to the perpendicular exchange field, as clearly demonstrated by our calculations. In addition, films in the Chern insulating phase display quasi one-dimensional (1D) spin-polarized gapless chiral edge states on the side-walls, the existence of which is ensured by the bulk-boundary correspondence. These chiral edge states are of course the ones responsible for the QAHE. There exist also non-chiral side-wall edge states, but these are gapped by ordinary space quantization in film with finite thickness and do not contribute to transport and therefore, do not ruin the QAHE if the films are thin enough \cite{pertsova2016,Pournaghavi_et_al_2021}.

In the Axion insulator phase, 1D gapless chiral edge states are absent and only non-chiral edge states, gapped if the film is thin, are present on the side walls. However, if the film is very thin, down to a few layers, a cross-over between a 3D magnetic TI to a 2D TI might occur. Crossovers from a 3D to a 2D TI occurring in thin films of Bi$_2$Se$_3$ were investigated in the early days of the work on TRS topological insulators such as BiSe and BiTe \cite{Zhand_NatPhys2010, Liu_PRB_2010}, and more recently also in experiments on even-layered thin films of intrinsic AFM TI MBT \cite{Lin_Nat_Comm_2022}. One would expect that for both cases, in the ultra-thin film limit, the 3D-2D crossover could result in a topological phase transition, which may display the characteristics of a 2D TI, namely the Quantum Spin Hall Effect (QSHE), which in our case is consistent with the calculated ${\cal C}_s$ = 1.

The occurrence of the QSHE should be accompanied by the presence of gapless counter-propagating helical edge states on the side-wall. This is indeed the case for ultra-thin films of TRS BiSe/Te TIs \cite{Zhand_NatPhys2010, Liu_PRB_2010}. For the case of intrinsic magnetic TI MBT ultra-thin films, the breaking of TRS by AFM magnetization causes the opening of an exchange gap in the band structure of side-wall helical edge states,
which is precisely what we find in our calculations, see Fig.~\ref{fig:fig1}(f), in agreement with the experimental results of Ref.~ \cite{Lin_Nat_Comm_2022}.

In Fig.~\ref{fig:fig1}(e-f) we present the results of the band-structure calculation of quasi-1D side-wall edge states of 110-surface of MBT thin films. For a 5-SL MBT thin film, we find chiral edge states, robust against disorder, propagating in one direction along the 1D boundary of the film, which is a distinct characteristic of the Chern insulating phase. 
% These unidirectional edge states are robust against magnetic perturbations with exchange field smaller than the 
% exchange gap opened in the top/bottom surface of odd-layered MBTs. 
Similarly for 6-SL films, we verified the presence of gapped side-wall edge helical states related to the QSHE, which can be viewed as a transport manifestation of the Axion insulating phase. Our findings support the notion that 5-SL and 6-SL MBT thin films are potential candidate systems to realize Chern and Axion insulating phases respectively. However, it is yet to be determined how surface and structural reconstructions affect the exchange gap and the topological character of these systems. Addressing this issue is the central part of this work.  

In Fig.~\ref{fig:fig2}(a), we take a 6-SL MBT film in its non-reconstructed form as a reference system having seven atomic-planes forming a 1T-like crystal structure. Considering only the bottom SL, the 1T-like atomic structure can transform in such a way that the Mn-atoms bonded to inner Te/Bi atoms form a 2H-type structure, as encircled in Fig.~\ref{fig:fig2}(b); we call this structure an interstitial-2H type reconstruction. It is important to note that such a structural reconstruction maintains the same stoichiometric ratio of atoms and does not alter the characteristic seven-planes of a MBT SL but rather introduces slight distortions in the structure. Alternatively, the bottom SL can reconstruct into a so-called peripheral-2H type, where the Bi/Te atoms at the edge form a 2H-type atomic configuration (see Fig.~\ref{fig:fig2}(e)). Because MBT thin films are obtained either from bulk samples via exfoliation techniques \cite{liu2020robust,ovchinnikov2021intertwined} or grown by Molecular Beam Epitaxy on various substrates at relatively higher temperatures (between 500$-$900 K) \cite{li2024imaging}, the possibility that such reconstructions occur in the synthesis process is highly likely, and so far it has been largely ignored. Therefore, we anticipate that different possible configurations can arise, e.g., either the top/bottom surface reconstruction shown in Fig.~\ref{fig:fig2}(c), or as other combinations as shown in Fig.~\ref{fig:fig2}(d-g). 

\begin{figure}[!t]
\includegraphics[width=0.5\textwidth]{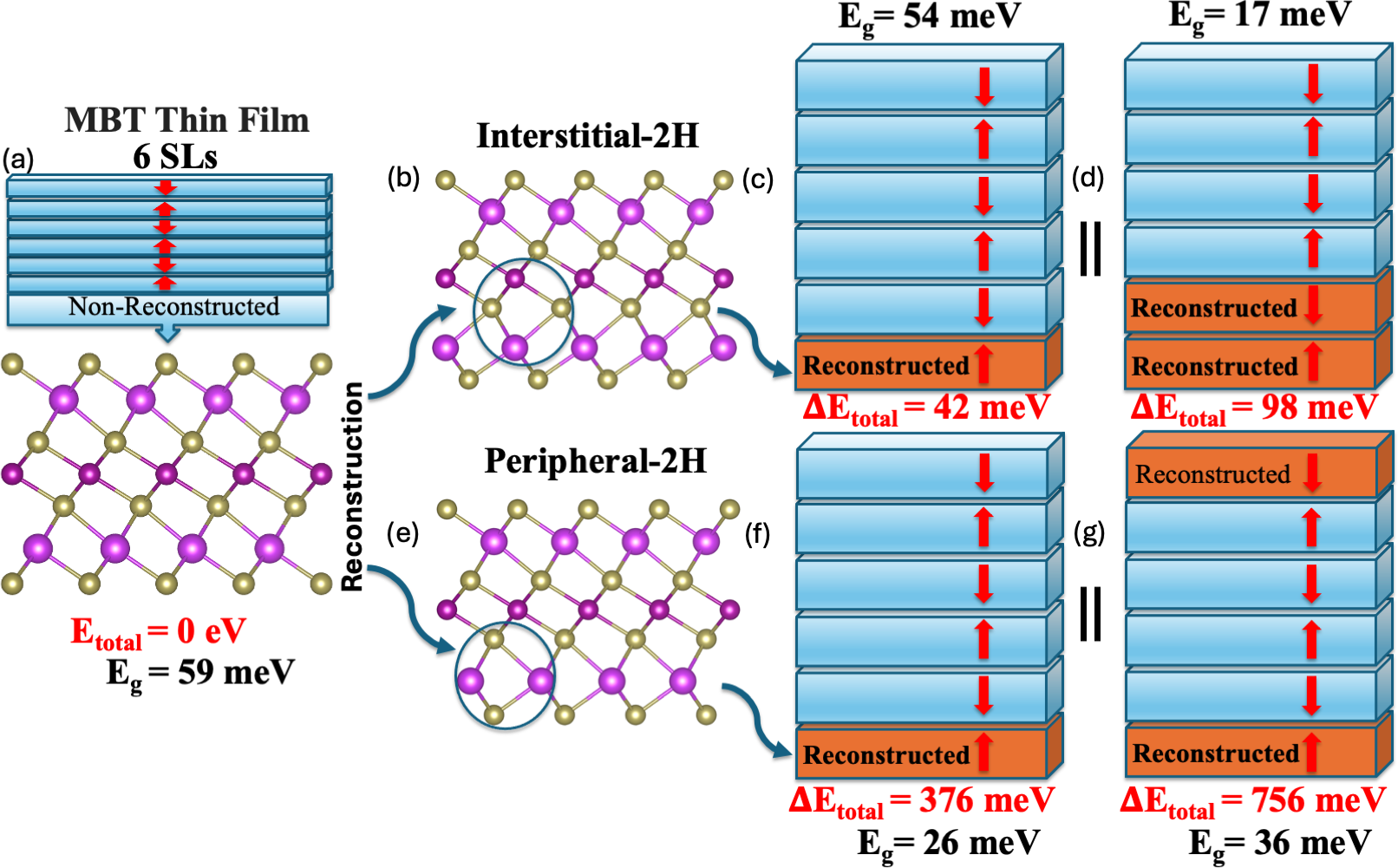}
\caption{(a) Schematic of a 6-SL MBT thin film in non-reconstructed form having an energy gap of $59$ meV, whose energy is taken as a reference (E$_{\text{total}}=0$). (b) The 1T-like structure transforms into a so-called interstitial-2H type reconstruction (encircled in blue) which manifests itself in different SLs in (c-d). Here, $\Delta$E$_{\text{total}}$ is the change in energy and new E$_g$ values are given explicitly. (e) A peripheral-2H reconstruction, with the encircled outer Bi/Te atoms making a 2H-structure. (f-g) display two possible combinations.}
\label{fig:fig2}
\end{figure}

To investigate this, we first looked at the energy landscape of these reconstructions. 
Defining $\Delta$E$_{\text{total}}$ as the energy difference between reconstructed (R) and non-reconstructed (NR) MBT structures after structural relaxation (i.e., $\Delta$E$_{\text{total}}$=E$_{\text{R}}$-E$_{\text{NR}}$), we found that the interstitial-2H structures are approximately 40-100 meV higher in energy compared to the unreconstructed one. In contrast to an existing study of Ref. \cite{HTan_BYan_PRL2023} on Mn-Bi antisite defective structures which are much higher in energy, these energy differences of interstitial-2H reconstructions to unreconstructed SL are rather small. On the other hand, for peripheral-2H type reconstructions, the energy differences are rather larger and amount to 376 meV and 752 meV, respectively, as shown in Fig.~\ref{fig:fig2}(f-g). Notably, for all reconstructions examined in this study, the energy difference between the R and NR cases remains unchanged for MBT thin films with thicknesses beyond 4-SL. Furthermore, we also performed nudged elastic band (NEB) calculations to determine energy barrier and reaction pathways between reconstructed and NR MBT thin films. Refer to Fig. S4-S5 in SI, a small energy barrier of 9.2 meV between peripheral-2H to interstitial-2H surface reconstruction indicates that latter is thermodynamically more stable than the former. Similarly, despite interstitial-2H reconstructed MBT surface in Fig. \ref{fig:fig2}(c) is only 40 meV higher in energy than the non-reconstructed MBT thin film, an energy barrier of 1.9 eV prohibits transformation of the former to the latter, as shown in
the Fig. S5. Since surface reconstructions in thin films are often driven by a combination of thermodynamic and kinetic factors, including surface energy minimization, substrate effect and strain relaxation, vacancies, and environmental conditions (e.g., temperature, pressure, and surface terminations), these reconstructions are likely to persist across different layer thicknesses.

In Fig.~\ref{fig:fig2}(c-d), we sketch possible interstitial-2H type reconstructions in a 6-SL MBT thin film. Due to atomic rearrangements in constituent SLs, we observe different top/bottom surface gaps starting from 54 meV (reconstruction only in the bottom SL) down to to 17 meV when reconstructions exist in two consecutive SLs. As shown in Fig.~S3, band inversion remains intact and robust in all cases. It is pertinent to mention that the combined $S=\Theta T_{1/2}$ symmetry in even-layered reconstructed MBT thin films is preserved, ensuring doubly degenerate bands corresponding only to the symmetric SL arrangement, as shown in Fig.~S3(a) of SI. Nevertheless, the systems remain in either the Axion or Chern insulating phase, depending on the parity number of SLs in the film, since the robustness of the band inversion guarantees the preservation of topological features.

\begin{figure}[!b]
\includegraphics[width=0.5\textwidth]{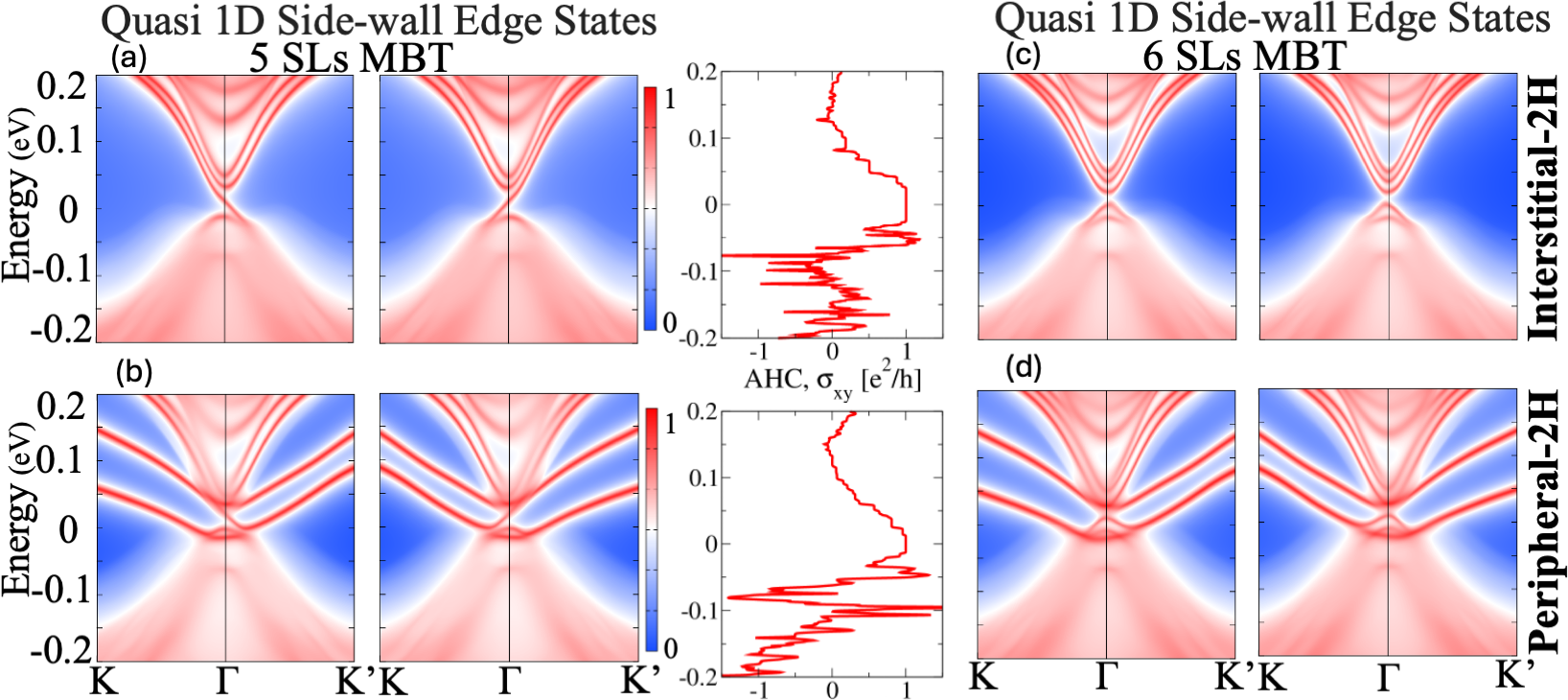}
\caption{(a-b) Quasi-1D side-wall edge states for interstitial-2H and peripheral-2H reconstructions in 5 SL MBT alongside anomalous Hall conductivity (AHC). (c-d) The same for 6 SL MBT thin film showing the appearance of additional surface states due to the peripheral-2H reconstructions.}
\label{fig:fig3}
\end{figure}

These observations are corroborated by the calculation of 1D Chiral edge states and anomalous Hall conductivity (AHC) for a 5-SL MBT film with interstitial-2H reconstructions, as shown below in Fig.~\ref{fig:fig3}(a). For the computation of the AHC, we used Kubo's formalism leading to the the following equation: 
\begin{equation}
    \sigma_{xy}^z=\frac{e^2}{\hbar}\sum_{n\neq n'} \frac{dk}{(2\pi)^2} [f(E_{n,k})-f(E_{n',k})]\Omega_{n,xy}^z (k), \label{equation:equation1}
\end{equation}
where 
\begin{equation*}
    \Omega_{n,xy}^z= -2\hbar^2 Im \sum_{\substack{n=occ,\\n'=unocc}} \frac{\langle u_{nk}\vert \hat v_y(k)\vert u_{n'k}\rangle\langle u_{n'k} \vert \hat{v}_x(k)\vert u_{nk} \rangle}{(E_{n,k}-E_{n',k})^2}
\end{equation*}
Here, $E_{n,k}$ is the eigenvalue of the bloch function $\vert u_{nk} \rangle$, $f(E_{n,k})$ is the Fermi-Dirac distribution function for a given chemical potential $\mu$. Besides, $\hat{v}_x$ and $\hat{v}_y$ are velocity operators.\newline 
While left and right chiral edge states persist under structural disorder, we note that AHC is quantized in units of e$^2/$h implying a (${\cal C}$)$=+1$ state for the 5 SL MBT film, as for the non-reconstructed case. The same is true for the 6-SL MBT thin-film, which maintains gapped top/bottom surfaces as well as side-wall edge states, as given in Fig.~\ref{fig:fig3}(c). Therefore, it is important to emphasize that interstitial-2H reconstructions can be differentiated from irremovable Mn-Bi antisite defects, which are known to severely perturb the topological character of the films due to the vanishing of the top/bottom surface gap \cite{wu2023irremovable, HTan_BYan_PRL2023}.
Moreover, this also explains the origin of a varying surface gap found in different studies while keeping the topological order intact for both even- and odd-layered systems.

\begin{figure}[!t]
\includegraphics[width=0.5\textwidth]{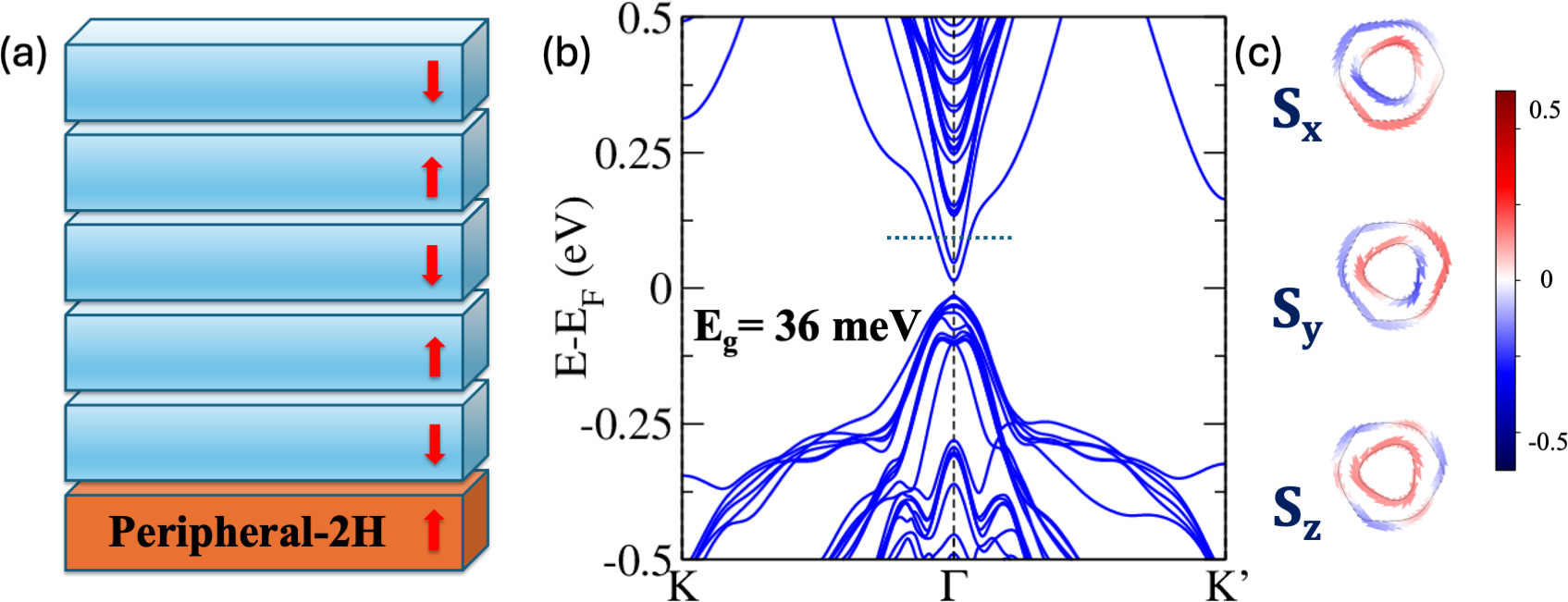}
\caption{(a) Schematic of the peripheral-2H reconstruction in 6-SL MBT thin film. (b) Electronic band-structure (BS) showing Rashba-like spin-split bands with E$_g$=36 meV. (c) 2D spin-textures projected on different spin-components for a fixed-energy cut at E=0.13 eV in BS, showing spin-momentum locking phenomena.}
\label{fig:fig4}
\end{figure}

Finally, we consider the case of peripheral-2H reconstructions which, although higher in energy compared to the non-reconstructed ground state, also displays surface gap modulation, as shown in Fig.~\ref{fig:fig2}(f-g). Therefore, we look only at the bottom surface reconstruction ($\Delta$E$_{\text{total}}$=376 meV) in both 5-SL and 6-SL MBT thin films, as shown in Fig.~\ref{fig:fig3}(b,d), respectively. Here, due to the presence of peripheral-2H reconstructions exactly at the edge of bottom SL, we observe additional states appearing close to Fermi energy interfering with existing edge-states. Nevertheless, despite a decrease in the top/bottom surface gap and the emergence of new states, the AHC remains quantized for 5-SL MBT. Hence, we conclude that the topological character in MBT thin films is robust against structural disorder and reconstructions (in the inner or outer SLs) as long as the top/bottom surface gap persists.

On the other hand, we observe that the peripheral-2H reconstruction (i.e., a reconstruction of Bi/Te atoms at the SL edge) induces large Rashba-like spin-splittings in the conduction bands of MBT thin films and displays spin-momentum locking as shown in Fig.~\ref{fig:fig4} for the case of a 6-SL MBT thin film. Analyzing the electronic dispersion of Fig.~\ref{fig:fig4}(b) and comparing to ARPES measurements in Ref. \cite{vidal2019surface,nevola2020coexistence} and others, we observe striking similarities, indicating that such splittings may arise due to edge SL reconstruction of peripheral-2H type. To confirm this, we plot 2D-spin textures projected on the in-plane and out-of-plane spin components corresponding to a fixed energy cut at E$=0.13$ eV in the band structure. Indeed, Fig.~\ref{fig:fig4}(c) shows that the spatial orientation of electron spins remains tangential to the $\Vec{k}$-vector. However, we note that peripheral-2H reconstruction may not be the only mechanism providing plausible explanation to the ARPES measurements. Alternative mechanisms, such as surface magnetism and stacking configurations, could also play a role. Furthermore, Refer to Fig. S6 in SI, a non-vanishing spin Hall conductivity for 6-SL peripheral-2H reconstructed thin film confirms the non-trival nature of this system.

\begin{figure}[!b]
\includegraphics[width=0.5\textwidth]{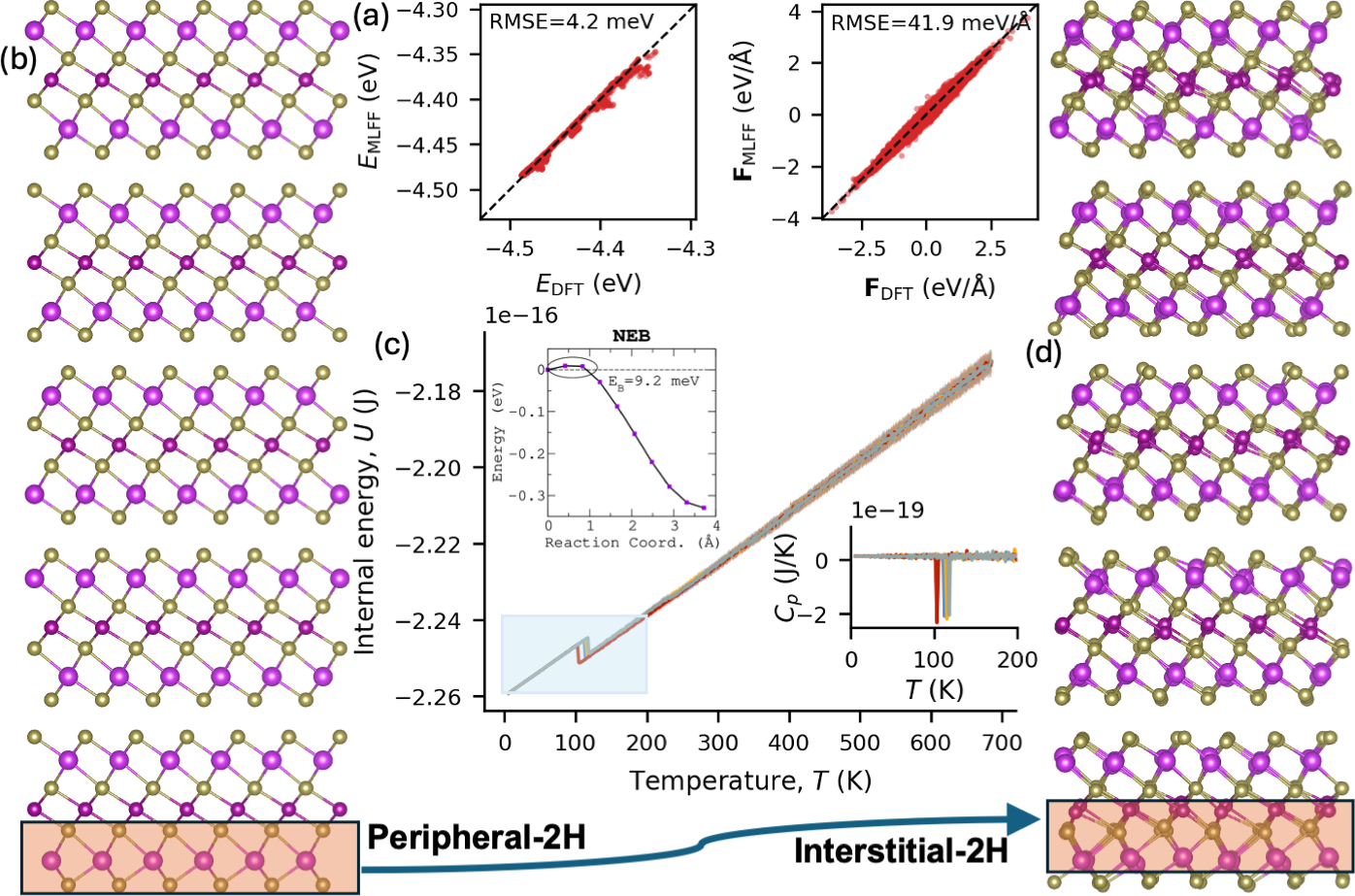}
\caption{(a) Regression plots in energy and force for the MLFF, here the root-mean-square-error (RMSE) is also shown. (b) The model system used for the MLFF driven MD simulations, a 5-SL MBT with peripheral-2H reconstruction in the bottom layer. (c) The internal energy as the system is heated from 5 to 700 K, here each colored line corresponds to an individual MD simulation. Top inset shows the minimum energy path between peripheral-2H and interstitial-2H surface reconstructions as obtained using nudged elastic band calculations. Bottom inset shows the specific heat capacity from 5 to 200 K with a clear phase transition seen around 100-110 K. (d) Snapshot of the final structure obtained from one of the MD simulations containing interstitial-2H reconstruction in the lower SL.}
\label{fig:fig5}
\end{figure}
To evaluate the thermal stability of the MBT surface reconstructions, we performed MD simulations driven by a MLFF. In the MD simulations, the system is heated from 5 to 700 K with a constant heating rate of 20 K/ns. The MLFF was trained on-the-fly using \textit{ab-initio} MD simulations and can predict both energies and forces with DFT levels of accuracy, as shown in Fig.~\ref{fig:fig5}(a). To ensure comprehensive sampling, we performed four independent MD simulations using an identical 5-SL MBT thin-film system. This model system, illustrated in Fig.~\ref{fig:fig5}(b), features a peripheral-2H reconstructed bottom surface. During the heating process, a phase transition from peripheral-2H to interstitial-2H occurs at a characteristic temperature of $112.51 \pm 5.61$ K. This transition is marked by a pronounced change in the system's internal energy, as shown in Fig.~\ref{fig:fig5}(c), as well as a corresponding peak in the specific heat, inset of Fig.~\ref{fig:fig5}(c). Post-transition, the interstitial-2H reconstruction exhibited remarkable stability, persisting over the remainder of the temperature range up to 700 K. The low transition temperature between these reconstructions can be attributed to the small energy barrier of 9.2 meV, as determined via nudged elastic band calculations, inset of Fig.~\ref{fig:fig5}(c). These results suggest that the peripheral-2H reconstruction is metastable, readily transitioning to the thermodynamically favored interstitial-2H configuration under thermal excitation. A snapshot of the final structure for one of the MD simulations shows the a MBT thin film consisting of interstitial-2H reconstructed bottom layer in Fig.~\ref{fig:fig5}(d). Note the atomic displacement in the upper layers due to the elevated temperatures. However, each SL preserves the characteristic 1T-like atomic geometries when reconstructions are not introduced among them.

\textbf{\textit{Conclusions}} -- 
We employed first-principles calculations and MLFF driven MD simulations driven to investigate surface reconstructions in Chern insulating and Axion-insulating MBT thin films. We demonstrate that interstitial-2H and peripheral-2H type reconstructions arising from the rearrangement of Mn and neighboring Bi/Te atoms are responsible for top/bottom surface gap modifications. While certain symmetries present in bulk do not remain intact in reconstructed MBT thin-films, we find that the topological character is robust against structural disorder and reconstructions both in the inner and outer regions of SLs. Our results for peripheral-2H reconstructions explain the origin of Rashba-like spin-split conduction band observed in ARPES experiment. This work provides a fundamental understanding of the occurrence and effect of disorder and surface reconstruction and its implications on the topological characteristics of MBT thin films.

%\section{Acknowledgements}
We thank S.~P.~Dash, A.~H. MacDonald, N.~Pournaghavi, A. Tyner, C.~Lei and P.~T.~Mahon for useful discussions. The work is financially supported by the Swedish Research Council (grant no: VR 2021-04622). The computations were enabled by resources provided by the National Academic Infrastructure for Supercomputing in Sweden (NAISS) partially funded by the Swedish Research Council through grant agreement no. 2022-06725 and the Centre for Scientific and Technical Computing at Lund University (LUNARC).

% The \nocite command causes all entries in a bibliography to be printed out
% whether or not they are actually referenced in the text. This is appropriate
% for the sample file to show the different styles of references, but authors
% most likely will not want to use it.
%\nocite{*}
\bibliographystyle{apsrev4-2}
\bibliography{main}% Produces the bibliography via BibTeX.

\end{document}